\documentclass{aa}
\usepackage{graphicx,txfonts,natbib,color}

\begin{document}
\title{The rotation and coma profiles of comet C/2004 Q2 (Machholz)\thanks{based on observations collected with the 1.2m Mercator telescope at the Roque de los Muchachos observatory, La Palma, Spain}}
   \subtitle{}
   \author{M. Reyniers\inst{1,2}\fnmsep\thanks{former Postdoctoral fellow of the
          Fund for Scientific Research, Flanders; now Scientific Assistant at
          Royal Meteorological Institute of Belgium (Belgian Federal Science
          Policy Office)}
          \and
           P. Degroote\inst{2}
          \and
           D. Bodewits\inst{3,4}
          \and
           J. Cuypers\inst{5}
          \and
           C. Waelkens\inst{2}
           }
   \offprints{M. Reyniers}
   \institute{Royal Meteorological Institute of Belgium, Observations
              Department, Ringlaan 3, 1180 Brussels, Belgium\\
         \email{maarten.reyniers@kmi-irm.be}
         \and
              Institute of Astronomy, Department of Physics and Astronomy,
              K.U.Leuven, Celestijnenlaan 200D, 3001 Leuven, Belgium\\
         \email{pieter.degroote@ster.kuleuven.be}
         \and
              KVI Atomic Physics, University of Groningen, Zernikelaan 25,
              9747AA Groningen, The Netherlands
         \and
              NASA Postdoctoral Fellow, Goddard Space Flight Center, Solar
              System Exploration Division, Mailstop 693, Greenbelt, MD\,20771,
              USA
         \email{dennis.bodewits@ssedmail.gsfc.nasa.gov}
         \and
              Royal Observatory of Belgium, Ringlaan 3, 1180 Brussels,
              Belgium\\
             \email{jan.cuypers@oma.be}
             }

   \date{Received 10 December 2007; accepted 14 November 2008}
   \authorrunning{M. Reyniers et al.}
   \titlerunning{The rotation and coma profiles of comet C/2004 Q2 (Machholz)}

  \abstract
  {}
  {Rotation periods of cometary nuclei are scarce, though important when studying
  the nature and origin of these objects. Our aim is to derive a rotation
  period for the nucleus of comet C/2004 Q2 (Machholz).}
  {C/2004 Q2 (Machholz) was monitored using the Merope CCD camera on the
  Mercator telescope at La Palma, Spain, in January 2005, during its closest
  approach to Earth, implying a high spatial resolution (50\,km per pixel).
  One hundred seventy images were recorded in three different photometric
  broadband filters, two blue ones (Geneva U and B) and one red (Cousins I).
  Magnitudes for the comet's optocentre were derived with very small apertures
  to isolate the contribution of the nucleus to the bright coma, including
  correction for the seeing. Our CCD photometry also permitted us to study the
  coma profile of the inner coma in the different bands.}
  {A rotation period for the nucleus of P\,=\,9.1$\pm$0.2\,h was derived. The
  period is on the short side compared to published periods of other comets,
  but still shorter periods are known. Nevertheless, comparing our results with
  images obtained in the narrowband CN filter, the possibility that our method
  sampled P/2 instead of P cannot be excluded. Coma profiles are also presented,
  and a terminal ejection velocity of the grains
  v$_{\rm gr}$\,=\,1609\,$\pm$\,48\,m\,s$^{-1}$ is found from the continuum
  profile in the I band.}
  {}
  \keywords{Comets: individual: C/2004 Q2 (Machholz)}

  \maketitle
%

\section{Introduction}
Comets are at the same time both the largest and the smallest celestial bodies
in our solar system. While the nucleus of a comet is typically 1 to 15\,km, the
ion tail can easily extend over more than 1\,AU \citep[e.g.][]{jones00}.
Cometary nuclei are chemically and morphologically complex bodies, consisting
of non-volatile ices and frozen grains. When these nuclei approach the sun,
gases start to sublimate from them, forming the cloud of gas and dust known as
the coma. For comets in the inner solar system, this coma is much brighter than
the nucleus, making the nucleus hidden from our view. Yet, information on the
dynamics of this nucleus is important to study not only the nature and origin
of these objects, but also more generally the origins of our solar system
itself. In the first place, it is of a statistical importance to determine
cometary periods, since a representative sample of periods could give important
clues on possible statistical dependencies with other properties, like orbital
period or mass, and with similar objects like asteroids. Moreover, information
on the rotational state of a comet gives important clues on the recent history
of the comet; fast rotation for example can point to strong recent activity in
which the comet changed its own rotation state, or alternatively (but less
likely) to a recent interaction with other solar system objects. Finally,
rotational periods are also an important parameter in the development of
sublimation models for nuclei \citep{schleicher98}.

\begin{table}
\caption{Basic properties of comet C/2004 Q2 (Machholz).}\label{tab:basicprprts}
\begin{center}
\begin{tabular}{ll}
\hline\hline
\multicolumn{2}{c}{\bf Comet C/2004 Q2 (Machholz)}\\
\hline
Discovery            & August 27, 2004\\
Earth passage    & $\Delta$\,=\,0.35\,AU on January 5, 2005\\
Perihelion & $r_h$\,=\,1.21\,AU on January 25, 2005\\
Orbital period               & $\sim$117000 year \\
Inclination & 38.6$^{\circ}$ with ecliptic\\
H$_2$O prod. rate & 2.6 10$^{29}$\,molec.\,s$^{-1}$ at $\Delta$\,=\,0.394\,AU ($\dag$)\\
\hline
\multicolumn{2}{l}{Refs.: Horizons On-Line Ephemeris System of JPL}\\
\multicolumn{2}{l}{(NASA) except $\dag$: \citealt{bonev06}}\\
\end{tabular}
\end{center}
\end{table}

In this paper, we present a study of the Oort Cloud comet C/2004 Q2 (Machholz).
Some basic properties of this comet are summarised in Table~\ref{tab:basicprprts}.
It was discovered on August 27, 2004 by Don E. Machholz \citep{machholz04},
and  reached a visual brightness of $\sim$3.5 mag. at its closest approach to
Earth, making it one of the most spectacular astronomical events of that year.
The observations presented in this paper are taken around this close encounter,
providing a high spatial resolution. In this paper, we concentrate both on the
nucleus, and on the inner (10$^5$\,km) coma. Our observational data consist
of CCD images in three photometric bands. Our first objective is to derive a
rotation period of the nucleus from our intensive photometric monitoring, while
our second goal is to study the profiles of the inner coma. These profiles are
relatively simple to determine, and directly lead to some interesting
properties, like the terminal ejection velocity of the dust grains.

Published rotation periods of cometary nuclei are scarce
\citep[see][for a summary]{samarasinha04}, since the nucleus of a comet
situated in the inner solar system is always hidden in an opaque and
overwhelming coma. Several techniques have been developed, each with their own
drawbacks. One possible technique is to search for structures in the inner coma
that emerge on the nucleus, like fans or jets, and search for periodicities in
their morphology. This technique was employed to search for the rotation state
of C/2004 Q2 (Machholz) by \citet{farnham07} and will be discussed later in this
paper (Sect.~\ref{subsect:rotltrtr}). The technique used in this paper is
aperture photometry using extremely small aperture radii. The idea is that in
the vicinity of the optocentre of the coma, a small but observable contribution
of the nucleus to the total brightness is expected. In the formalism of
\citet{licandro00}, if the brightness measured in an aperture $\rho$ around the
optocentre is given by
\begin{displaymath}
B(\rho) = B_N + B_C(\rho)
\end{displaymath}
where $B_C(\rho)$ is the brightness of the coma captured in a radius $\rho$ and
$B_N$ the nucleus brightness. The amplitude of the light curve is then
\begin{equation}
\Delta m (\rho) = -2.5 \log [{\rm min}(B_{N}) + B_C(\rho)] / [{\rm max}(B_{N}) + B_C(\rho)]. \label{eq:licandroqtn}
\end{equation}
Note that in this equation, we implicitly assume that $B_C(\rho)$ is
constant during the time span of the observations. In practice, changes in
the activity can imply large coma variabilities, especially near perihelion.
Nevertheless, from Eq. (\ref{eq:licandroqtn}), it is easily seen that when taking large apertures, the contribution of the
coma $B_C(\rho)$ cancels out the amplitude of the light curve induced by the
variability of the nucleus. Therefore, very small apertures around the comet's
optocentre are usually employed to detect variability induced by the nucleus
\citep{jewitt85, jewitt90, meech93, meech97}.
However, a problem arises when taking apertures smaller than the seeing disc
of the observation. This problem was studied in detail by \citet{licandro00},
who dubbed this problem the {\em seeing effect}. They found that several
published periods were in fact contaminated, or even attributable, to this
seeing effect. In our study, we will take this effect into account.

\begin{figure*}
\begin{center}
\resizebox{16 cm}{!}{\rotatebox{-90}{\includegraphics{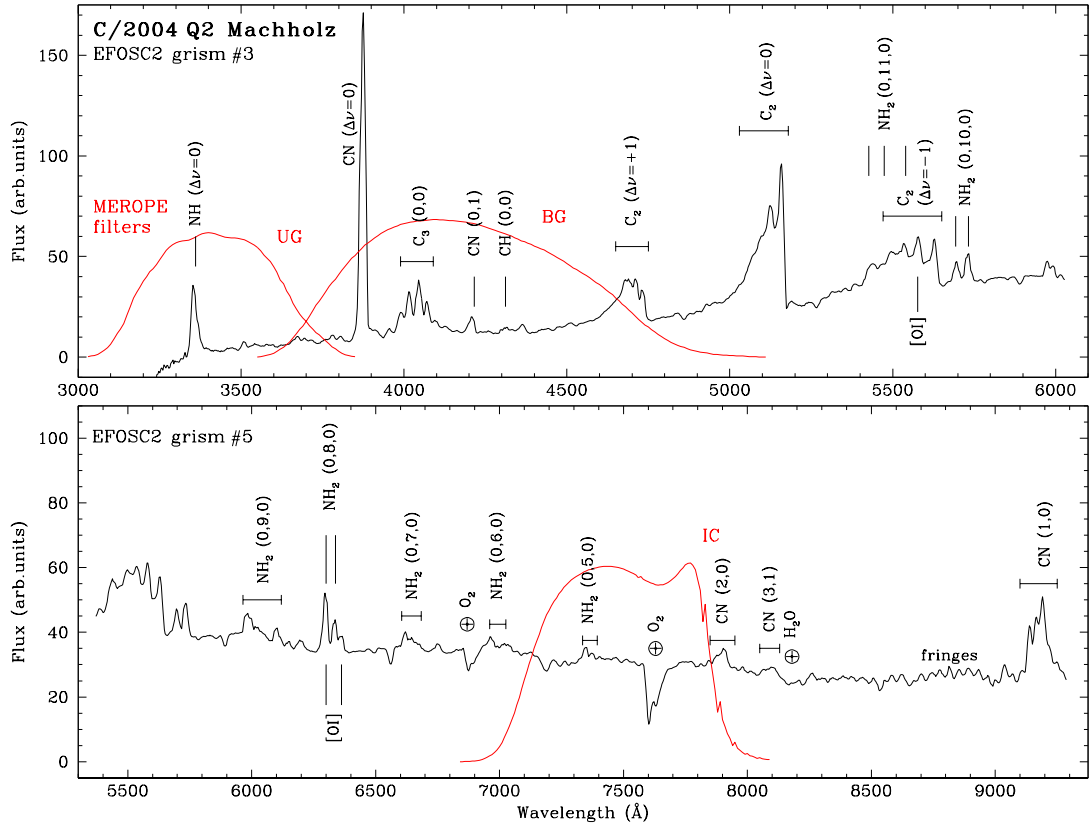}}}
\end{center} \vskip -.8 cm
\caption{The low-resolution spectrum of comet C/2004 Q2 (Machholz) taken with
the EFOSC2 instrument taken with the ESO 3.6\,m at La Silla on 9 december 2004.
See the text for the line identification. The passbands of the three photometric
filters of the Merope instrument used in this paper are shown in red. It is
obvious that these broad filters cover a mix of continuum and emission
features, with the I filter having the most emission-free passband. The two
spectra are ESO archive data, and are taken under programme 274.C-5020
(P.I. Weiler).}\label{fig:lowrsltn}
\end{figure*}

The reduction of the data, as well as its correction for the seeing effect,
are an important part of our work, since it will ultimately determine the
quality of our analysis. Therefore, the technical part of our analysis will
be a considerable fraction of this paper: The observations are described in 
Sect.~\ref{sect:observtns}, while the reduction is discussed in
Sect.~\ref{sect:reductn}; Sect.~\ref{sect:extrctn} is devoted to the extraction
of the cometary magnitudes, including the correction for the seeing effect;
Next, the variability analysis and its results are presented in
Sect.~\ref{sect:variabilitynlss}; In Sect.~\ref{sect:comaprfls} we construct
the coma profiles in the three filters that we used for this programme.
The general discussion of the obtained results is found in
Sect.~\ref{sect:generaldscssn}; We end with the conclusions
(Sect.~\ref{sect:finalcnclsns}).

\begin{table}
\caption{Observational log. For each night, the number of observations is
given for each filter U, B and I. Also the heliocentric ($r_h$) and geocentric
($\Delta$) distances are given (in AU, at 0:00\,UT), together with the
correction factor $f_a$ (see Eqn. \ref{eqn:fa} in Sect.~\ref{sect:distancecrrctn}).}\label{tab:observationallg}
\begin{tabular}{ccccccc}
\hline\hline
\multicolumn{7}{c}{Observational log}\\
\hline
            & n(U) & n(B) & n(I) & $r_h$ & $\Delta$ & $f_a$\\
\hline
2005-Jan-08 & 5 & 5 & 5 &1.2337 & 0.3485 & 1.00\\
2005-Jan-09 & 9 & 9 & 9 &1.2305 & 0.3500 & 1.00\\
2005-Jan-10 &10& 9 & 8 &1.2274 & 0.3520 & 1.01\\
2005-Jan-11 & 8 & 9 & 8 &1.2245 & 0.3546 & 1.02\\
2005-Jan-12 & 8 & 8 & 8 &1.2219 & 0.3577 & 1.03\\
2005-Jan-13 & 8 & 8 & 8 &1.2194 & 0.3613 & 1.05\\
2005-Jan-14 & 6 & 6 & 6 &1.2171 & 0.3654 & 1.07\\
2005-Jan-15 & 5 & 5 & 5 &1.2150 & 0.3699 & 1.09\\
\hline
\end{tabular}
\end{table}

\section{Observations}\label{sect:observtns}
Comet C/2004 Q2 (Machholz) was intensively monitored with the 1.2\,m Mercator
telescope at Roque de los Muchachos at La Palma (Spain), between January 7 and
January 15, 2005. The telescope is equipped with a photometric camera
{\em Merope} \citep{davignon04}, consisting of a 2k$\times$2k EEV CCD with a
filter wheel consisting of the traditional 7-band Geneva system
\citep{golay66, rufener88}, added with a Gunn R \citep{thuan76} and a Cousins I
\citep{cousins75} filter. The field of view is a square 6.\arcmin5, resulting in
a 0.\arcsec19 sampling.

\begin{figure}
\begin{center}
\resizebox{\hsize}{!}{\includegraphics{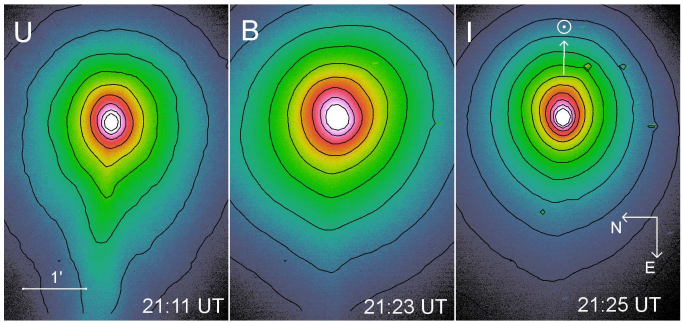}}
\end{center}
\caption{Three observations, one in each band, taken on January 11, 2005. The
frames are shown here to give an idea of the type of data. They are taken
consecutively, and hence under very similar ambient conditions. In the U
filter, the very clear signature of the ion tail is visible, in B this tail is
only marginally visible, and in I it is absent. The field of view of these
images is 3.\arcmin6\,$\times$\,5.\arcmin0 of the total square 6.\arcmin5 of the
CCD; the scale is shown in the lower left corner. Note that on these images,
the core of the coma seems to be saturated, which is obviously not the case,
but solely a consequence of the chosen colour scale.}\label{fig:panelps}
\end{figure}

181 Images were taken, in three different bands: 61 observations in both the
Geneva U and B band, and 59 in the Cousins I band. A log of the observations is
given in Table~\ref{tab:observationallg}, while three typical observations are
shown in Fig.~\ref{fig:panelps}. Integration times were
fixed at 60 sec for B and I exposures, and 600 sec for U band exposures.
Each set of observations was accompanied with a standard star observation,
immediately before, after or inbetween the comet observations. Standard
stars were taken from the Geneva catalogue of standard stars \citep{rufener99}.
Special care was taken to ensure that the standard stars were not situated in
the extended coma or tail of the comet. Standard stars are used to accurately
determine the extinction conditions of the time of observation. In addition, 40
sky flats in the three used filters were taken in this observing run, 16 in U,
11 in B and 13 in I. Few comet observations ($\sim$10) could not be used for
further analysis, mostly due to a wrong tracking or pointing of the telescope.
Weather conditions were exceptionally good with clear nights and a superb seeing
throughout the whole observing run.

Before we continue in discussing the actual reduction of the CCD frames,
it is a very useful exercise to get an idea which cometary emission these
three filters include. For this purpose, we downloaded low-resolution optical
spectra of C/2004 Q2 (Machholz) from the ESO archive
({\tt\small http://archive.eso.org}). These archive data are taken on December
9, 2004, with the EFOSC2 instrument that is mounted on the ESO 3.6m telescope,
under observation programme 274.C-5020 (P.I. Weiler). Reduction was done in the
available ESO-MIDAS routines by ourselves, and included the standard steps of
a long-slit reduction. Extraction was done over the whole spatial direction. We
succeeded to identify all major emission features in these spectra, taking
laboratory wavelengths from different sources
\citep{kawakita02, lara04, odell71}. High resolution atlases of
cometary emission lines \citep[e.g.][]{cochran02} indicate that many more
emission features are present, but these are not resolved at this relatively
low resolution. From a simple inspection of this figure, it is clear that the
Geneva U and B bands are sampling both continuum radiation and emission
features: NH (and most probably also the very strong OH emission at
3085\,\AA) in the U band; in the B band the strong CN\,($\Delta$$\nu$\,=\,0)
feature, but also C$_3$ and CH features. The I filter, however, does not contain
such strong emission features, and mainly samples the dust continuum.

\section{Reduction}\label{sect:reductn}
The reduction of CCD frames involves the bias correction, the construction of a
master flat field, the correction of the raw science frames by this master flat
field, and the correction for the background.

Bias correction was done by using the pre- and overscan regions of the image.
It was not possible to construct one master flat per filter for the entire
observing run, since a few dust particles situated in front of the CCD were
clearly visible on the flats. Therefore, several master flats were created per
filter, each of them constructed with raw flats that show the same pattern of
dust particles. Flats were combined by a median and sigma-clipping method,
while special attention was drawn to an adequate removal of the few weak stars
present in some raw flats. Note that none of the pixels that were affected
by the removal of those weak stars, coincide with pixels that were used in
the extraction of the comet's optocentre (Sect.~\ref{subsect:cometxtrctn}).

The background correction was also not straightforward. In the case of the
standard stars, a clear gradient is present in the background, probably caused
by the electronics involved in the windowed (300$\times$300 pixels) read-out
of these frames. The gradient was, however, efficiently removed by the
construction and subtraction of a second degree polynomial. In the case of
the comet frames, the background could not be determined on the frame, since
the coma of the comet is much larger than the field of view of our detector. No
dedicated sky frames were taken during the observing run. Consequently, no
background correction was done on the comet frames. This choice can be
motivated by two arguments. First, a rough estimate of the sky contribution
could be made on science frames of other science programmes that night, and we
concluded that on all frames the sky contribution is several orders of
magnitude weaker than the coma, even on the borders of the CCD. Second, if we
fit the decline of the coma with a polynomial f($\rho$) of $n$th degree,
starting from the centre of the coma, then the limit of f($\rho$) for
$\rho$$\rightarrow$$\infty$ is a useable estimate of the background
contribution \citep[e.g.][]{schulz95}. In all our fits, this limit $a_0$ is
significantly small, although there is some difference in $a_0$ for different
polynomial degrees, and for different fitting directions away from the centre.
Moreover, the observations were carried out around new moon (January 10),
assuring a minimal sky contribution. In any case, the background does certainly
not exceed a few thousandths of the pixel values around the centre of the coma.
Therefore, background subtraction can be safely neglected when concentrating on
the central luminosity, as we do in this paper.

\begin{figure}
\resizebox{\hsize}{!}{\includegraphics{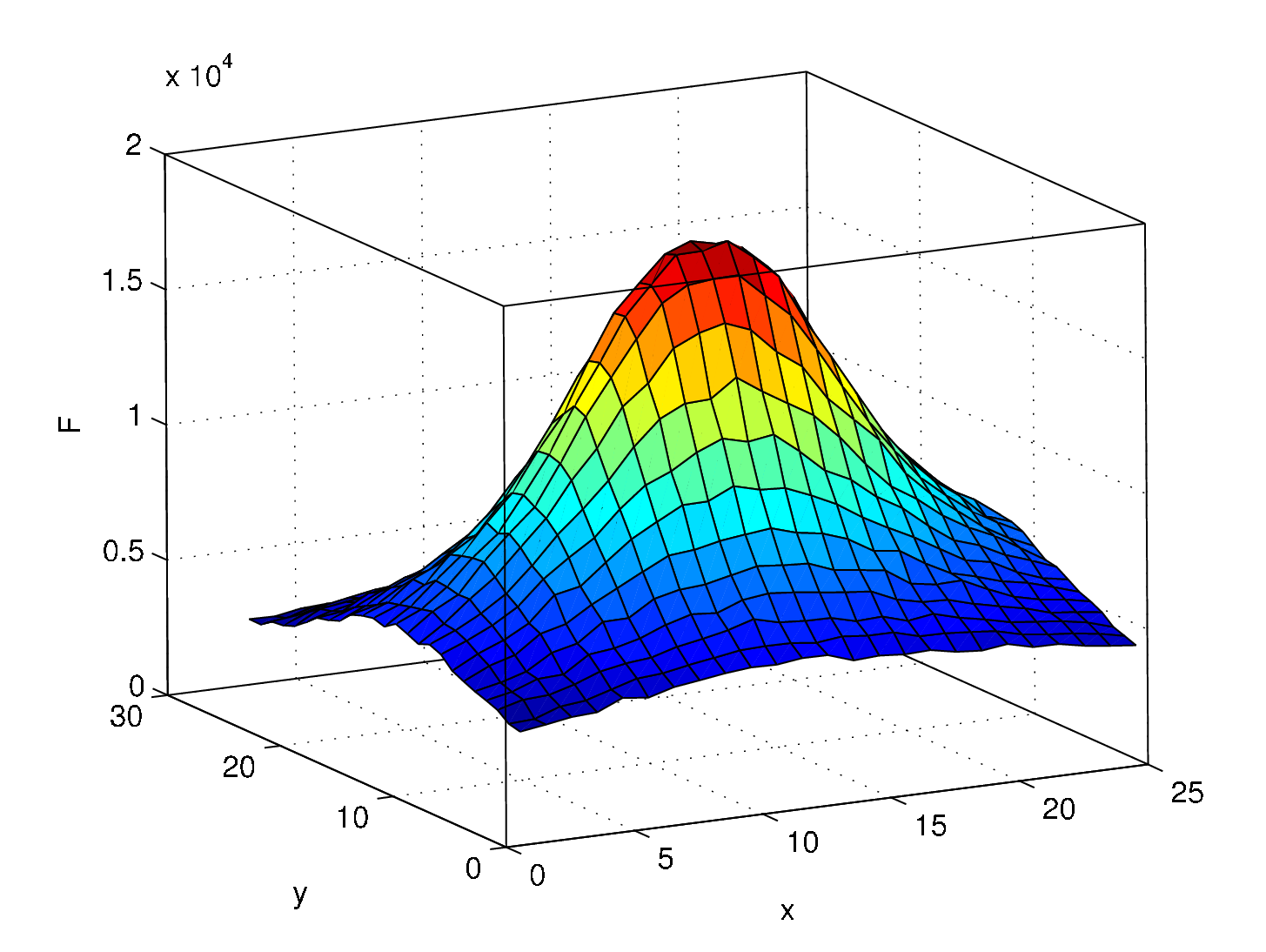}\includegraphics{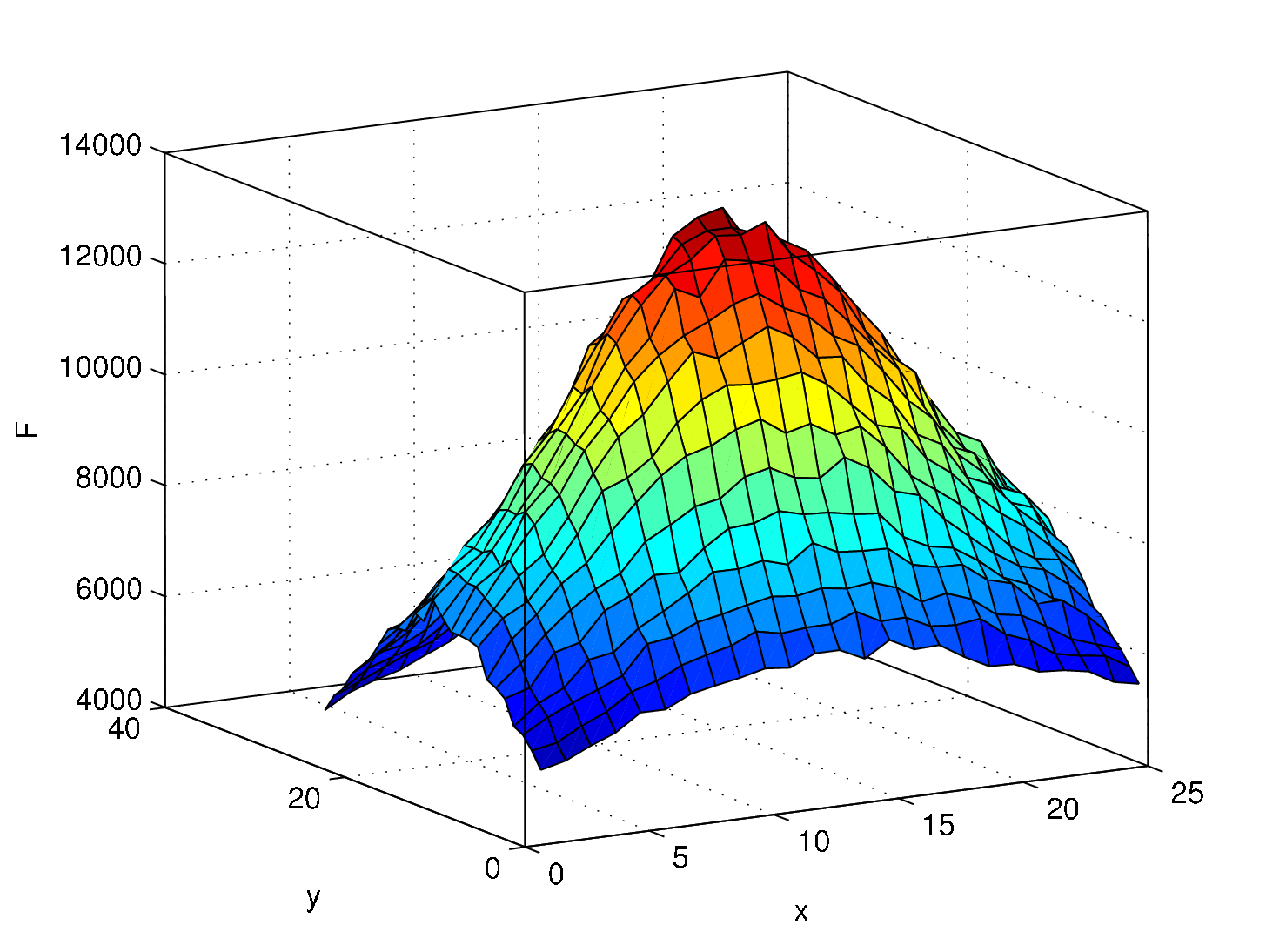}}\\
\resizebox{\hsize}{!}{\includegraphics{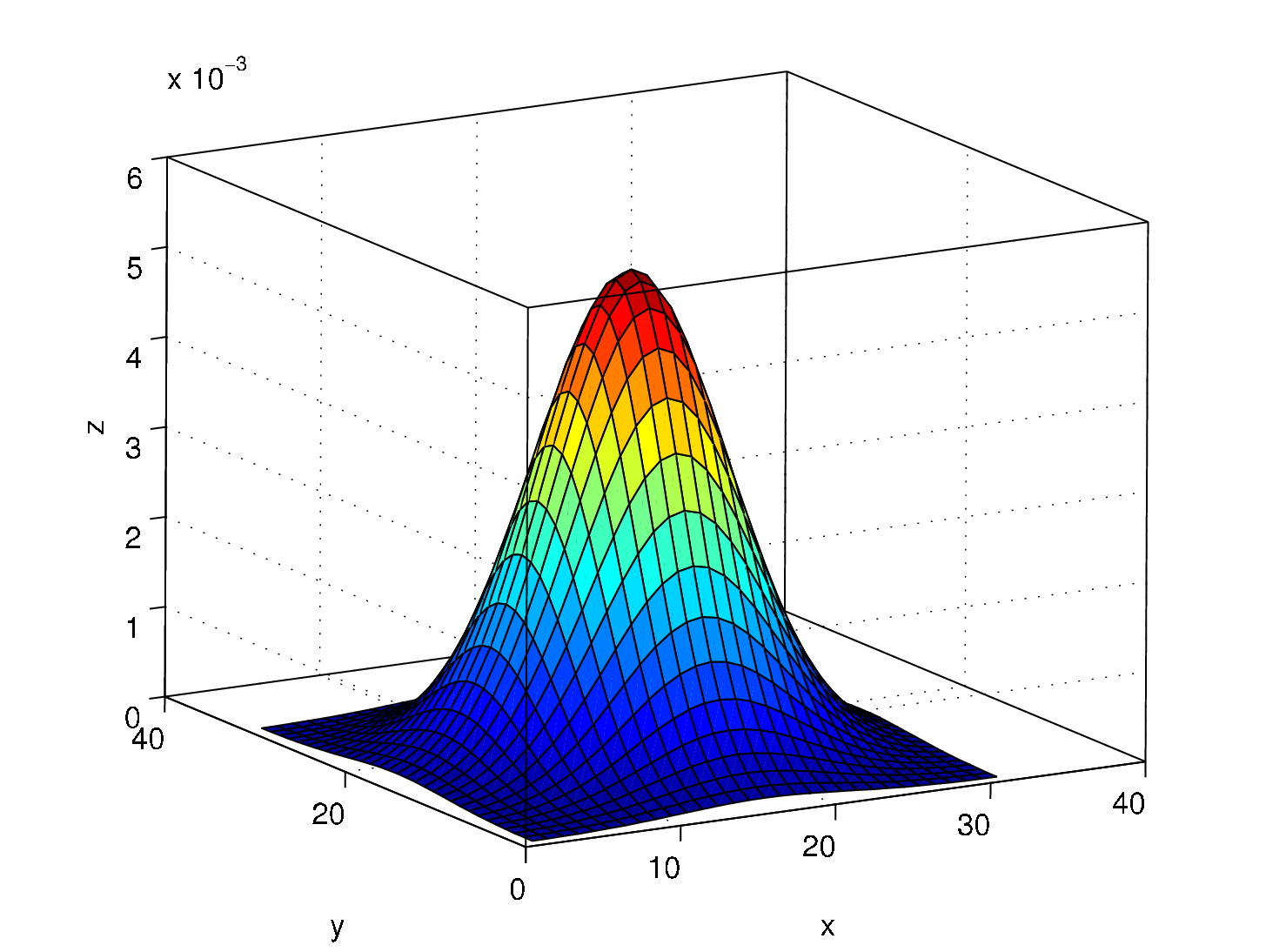}\includegraphics{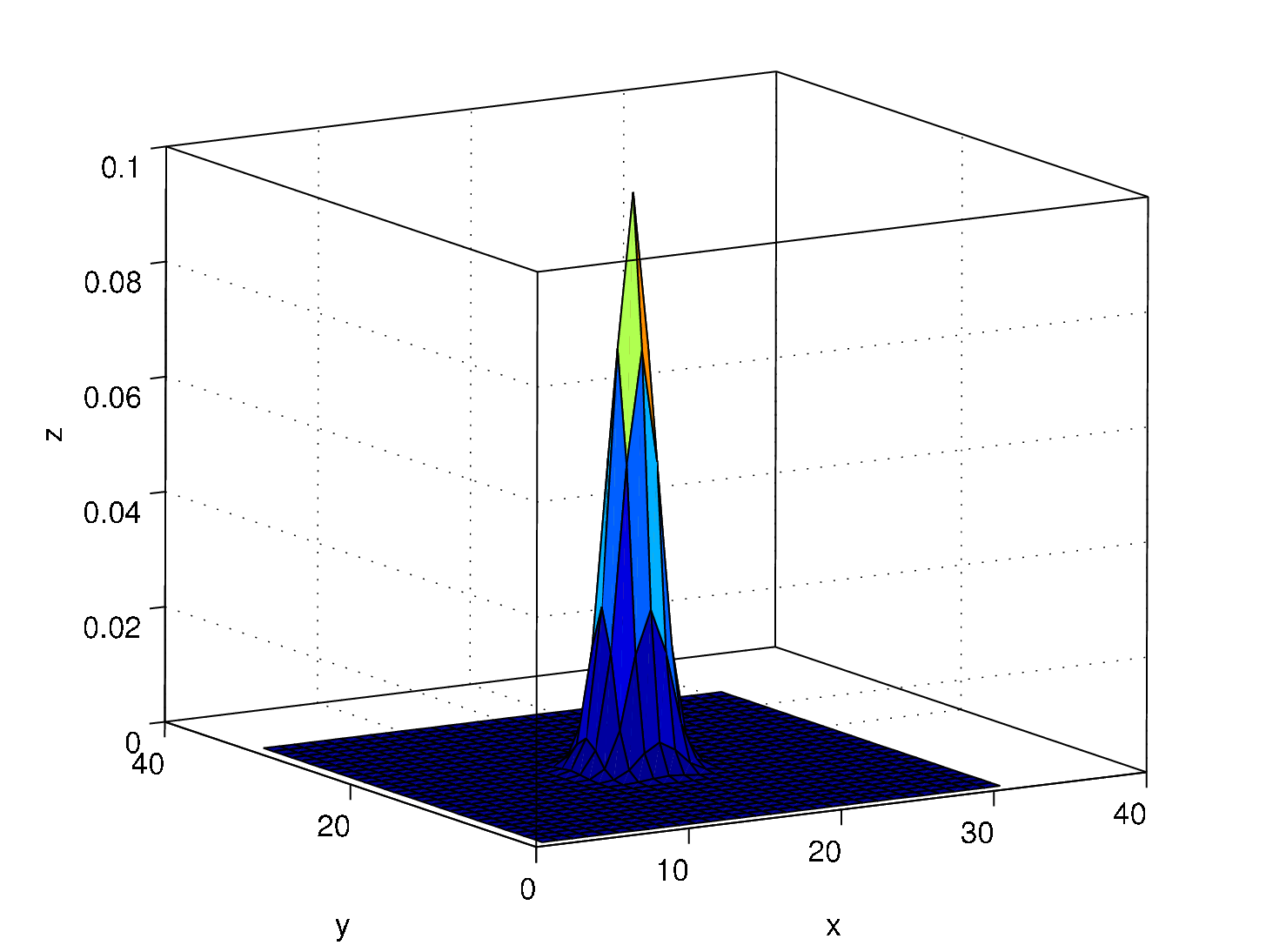}}\\
\resizebox{\hsize}{!}{\includegraphics{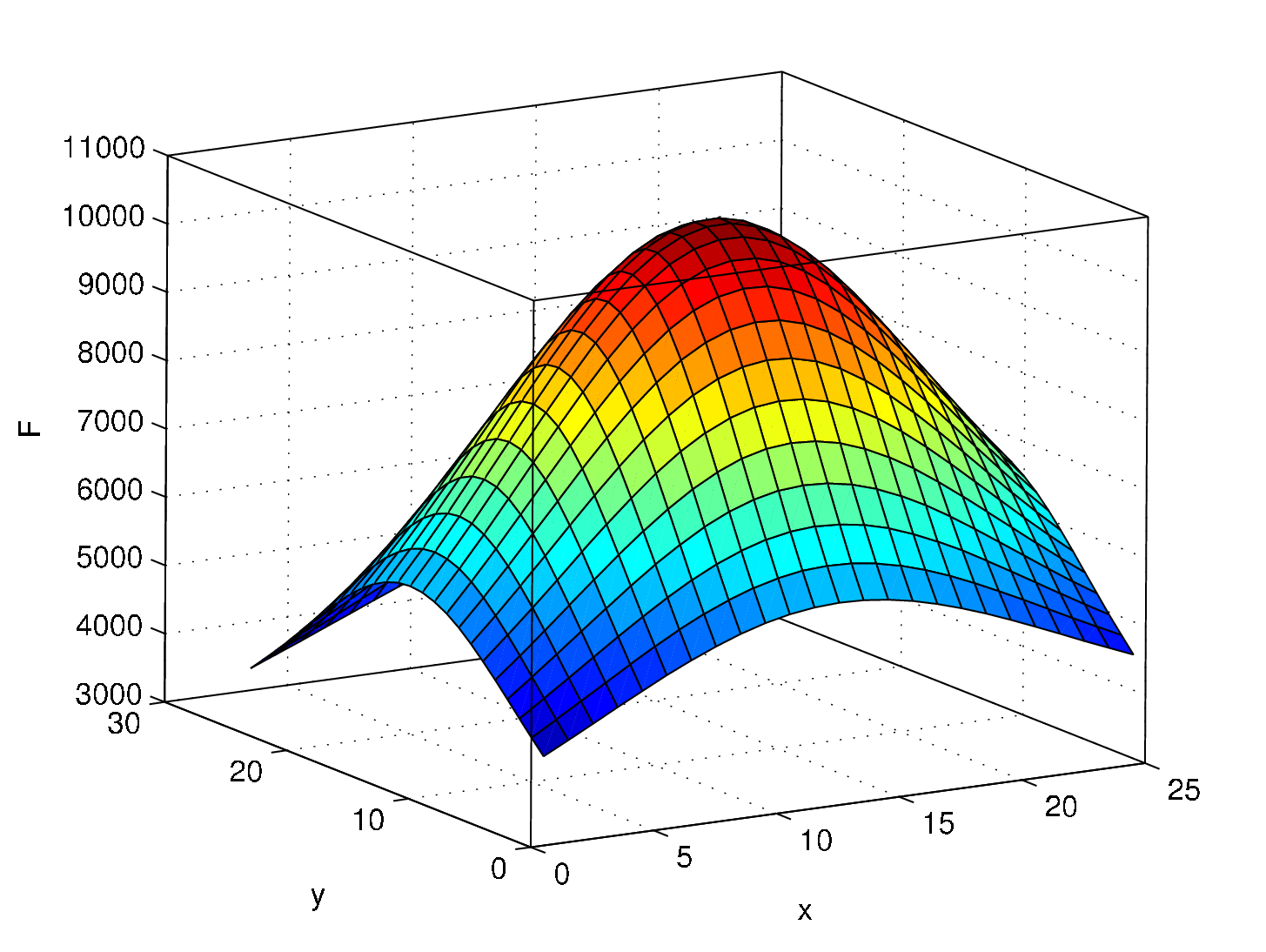}\includegraphics{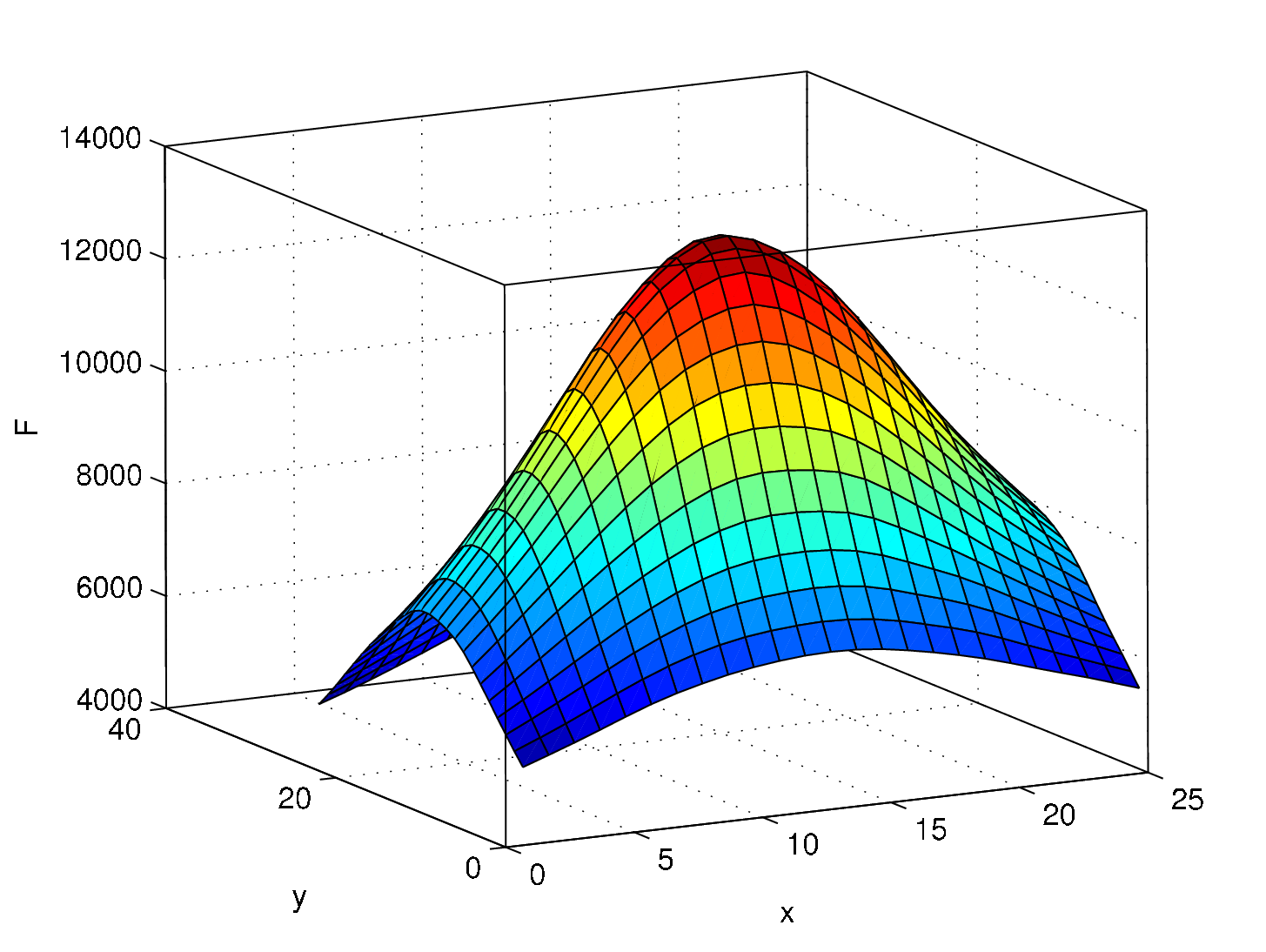}}
\caption{Correction for the {\em seeing effect} for a frame taken in optimal
conditions (left column of figures), and for a frame taken in less favourable
conditions (right column). The upper row shows the optocentres of the reduced
frames (but not yet seeing-corrected). To degrade these frames to the same
seeing, the left one has to be convolved with a broad Gaussian kernel (middle
left panel), while the right one with a very narrow Gaussian kernel (middle
right panel). The resulting optocentres are show in the bottom panels.}\label{fig:twokrnks}
\end{figure}

\section{Extraction}\label{sect:extrctn}
\subsection{Extraction of standard stars}
Due to their brightness, standard stars are observed out of focus. Standard
stars are extracted using SourceExtractor \citep{bertin96}. The method of
extraction is a simple addition of the flux above a certain threshold
(1.4$\sigma$), with a Gaussian correction for the wings of the star.
From these standard star magnitudes, extinction coefficients $k_{\lambda}$
are derived in each filter. These extinction values were then interpolated
to the observation times of the comet.

The extinction values in the U and B filters correlate very well, which is
an independent check for their accuracy. The mean $\overline{k_{\lambda}}$,
the range (1\,$\sigma$) and the peak-to-peak of the extinction values in the U
filter for the whole observing period are as follows: 
$\overline{k_{\lambda}}$\,=\,0.55\,$\pm$\,0.03 with peak-to-peak 0.12. For the
B filter, these values are $\overline{k_{\lambda}}$\,=\,0.30\,$\pm$\,0.02 and
peak-to-peak 0.07. These basic statistics show that the extinction values are
very stable throughout the observing run. We have also performed a frequency
analysis of the extinction values for these two filters, but no significant
frequencies were found in the range [0,\,10]\,c\,d$^{-1}$.

The extinction values in the I filter are much more variable than the values
for U and B. Moreover, they do not correlate with the two other filters. Since
we could not find an explanation for this result, we chose to derive the
extinction in the I band from a linear combination of the U and B band
extinctions. Luckily, the extinction correction is not so critical for the I
band, since it is generally very small ($\la$0.1\,mag).

\subsection{Distance correction of the comet frames}\label{sect:distancecrrctn}
Before the actual extraction of the coma centre, one has to correct for the
motion of the comet both relative to the observer and the sun. During our
observing run, both effects have an opposite effect, as the comet approaches
the sun while it recedes from the earth. To correct for the luminosity change
caused by the motion of the comet, we have multiplied the science frames
by
\begin{equation}
f_a = \left(\frac{\Delta}{\Delta_0}\right)^2 \left(\frac{r_h}{r_{h_0}}\right)^2 \label{eqn:fa}
\end{equation}
with $\Delta$ the distance comet-earth, and $r_h$ the distance comet-sun,
whereas the subscript $0$ denotes these distances on  the first night of our
observations. The values of these quantities were retrieved from the Horizons
On-Line Ephemeris System from the Solar System Dynamics Group of JPL (NASA)
({\tt\small http://ssd.jpl.nasa.gov/horizons.cgi}) and are given in
Table~\ref{tab:observationallg}. The factor $f_a$ reaches a value of 1.09 in the
last night, resulting in a magnitude decrease of $\sim$0.04.

\subsection{Correction for the seeing effect}\label{sect:seeingcrrctn}
In this particular case, especially the extraction of the science frames (see
Sect.~\ref{subsect:cometxtrctn}) is a delicate step, since we will deal with
extremely small aperture sizes, much smaller than the seeing disc. Since seeing
conditions are expected to be variable throughout the night, extraction with a
aperture size smaller than a typical seeing will introduce artificial
variability. In a paper devoted to this {\em seeing effect}, \citet{licandro00}
claim that several published rotational periods of comets are doubtful, and
are probably affected by this seeing effect. It is therefore extremely
important to account for the variation in the seeing.

\begin{figure}
\resizebox{\hsize}{!}{\rotatebox{-90}{\includegraphics{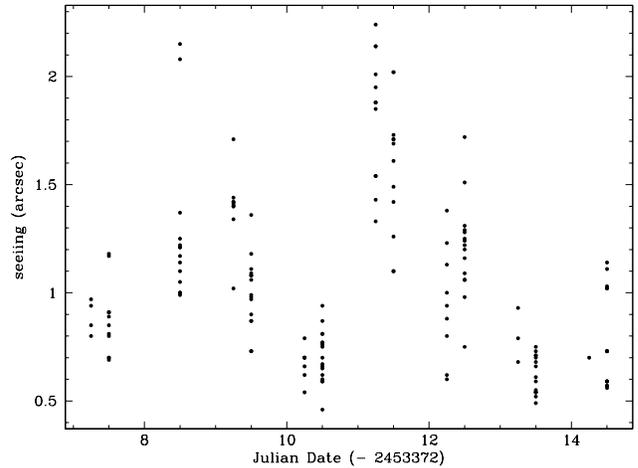}}}
\caption{The evolution of the seeing during our observational run as recorded
by the RoboDIMM telescope \citep{augusteijn01, omahony03}, interpolated to the
times of our comet observations.}\label{fig:robodimmsng}
\end{figure}

Detailed and accurate seeing measurements are available on the Roque de los
Muchachos site, provided by the dedicated RoboDIMM telescope 
\citep{augusteijn01, omahony03} of the ING. RoboDIMM forms four images of
the same star, measuring image motion in two orthogonal directions from each of
the two pairs of images, from which it derives four simultaneous and independent
estimates of the seeing. The values are automatically corrected for the zenith
distance and a wavelength of 550nm. The time resolution of these measurements
is 2 minutes, and therefore very well suited for our project. We will use these
data for an adequate correction of the seeing effect. The idea is to degrade the
seeing of each science frame to the worst seeing that was recorded during the
comet observations. While this method seems to be quite rough, its application
can be motivated by two facts. First, thanks to the good seeing conditions, the
``worst seeing'' was not that bad at all. In Fig \ref{fig:robodimmsng}, the
RoboDIMM seeing values interpolated to the comet observation times are shown.
It is clear that the least favourable seeing does not exceed a FWHM of
$\sim$2.\arcsec2. Second, the reverse process, in which frames taken in bad
seeing conditions are deconvolved to an average or a very good seeing, is a
much less robust process. The comets frames were degraded to the worst seeing
in that filter, by convolution of a twodimensional Gaussian kernel. As a
consequence, images taken in good seeing conditions are convolved with a broad
kernel, while images taken in worse conditions, only need a small kernel to be
degraded with. On Fig.~\ref{fig:twokrnks}, two examples are show of such a
convolution. The left column shows the technique for a frame taken in good
conditions, while the right column shows the same set of frames for a frame
taken in bad conditions. From top to bottom this figure shows: the frames as
they come out from the reduction (upper row), the Gaussian kernel to convolve
with (middle row), and the resulting, convolved frames (lower row). The upper
left frame shows the optocentre of a comet frame taken at good conditions.
Consequently it needs quite a broad kernel (middle left frame) to degrade it to
desired seeing. The reverse applies to the right set of frames, with the raw
frame taken in less good conditions.

Note that an objection could be made against the use of seeing data
originating from another source than our own telescope. Indeed, the seeing is
computed for a region of the sky that is not necessary the same as the region
containing the comet. Nevertheless, we chose this option for several reasons.
(i) A large part of the observing run was photometric, and hence only small
spatial variations of the seeing value are expected; (ii) Our own observational
data do not permit to deduce a seeing value: the standard stars are taken out
of focus (to prevent saturation), the science frames are tracked on the comet,
and no other science frames are available inbetween the comet observations.

\subsection{Extraction of the comet}\label{subsect:cometxtrctn}
Now that all comet frames are degraded to the same seeing, we can safely
extract the optocentre of the comet with small apertures well below the seeing
disc. This technique poses, on its own turn, two specific problems to deal
with. The first one is the determination of the exact location of the
optocentre, while the second one involves the discretisation of the pixels when
using very small apertures. The extraction is then done using three different
extraction radii (1, 3 and 6 pixels).

\subsubsection{Determination of the centre}
Several solutions can be put forward to determine the optocentre of the coma.
The most simple one is obviously taking the position of the brightest pixel
in the coma. A second one is to fit a twodimensional Gaussian in the proximity
of the optocentre. A third possibility is to take the barycentre of a limited
region around the optocentre. We chose this last method, since some tests
showed that the first method was a too rough determination, while
the Gaussian method did not give good results either. The Gaussian method
obviously assumes a Gaussian profile for the optocentre. While this is indeed
the case for the immediate surroundings of the optocentre due to the effect of
the seeing, it is not the case for a larger area since the coma shows a profile
different from Gaussian (see Sect.~\ref{sect:comaprfls}).

\subsubsection{Discretisation of pixels}
In the case of extraction with small apertures, flux has to be added from the
sub-pixel level. Common extraction programs, such as SExtractor, divide each
pixel in a certain number of virtual sub-pixels (5$\times$5 in the case of
SExtractor), and add the flux in those sub-pixels lying in the predefined
aperture radius. Here we chose to work strictly analytic. For each pixel lying
at the border of the aperture radius, we calculated the exact fraction of that
pixel in the radius.

\section{Variability analysis}\label{sect:variabilitynlss}
\begin{figure}
\resizebox{\hsize}{!}{\includegraphics{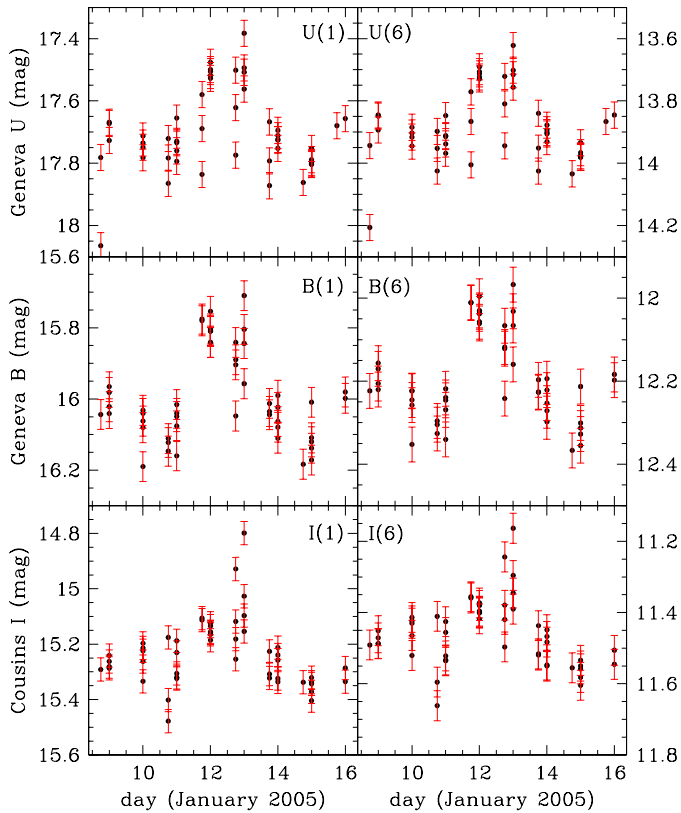}}
\caption{Overview of the magnitudes for the extracted optocentre of the comet
in the three different bands. Shown here are the results for an aperture
radius of 1 and 6 pixels (0.\arcsec19 and 1.\arcsec14).}\label{fig:alledt}
\end{figure}

The final magnitudes for the optocentre of the comet are graphically presented
in Fig.~\ref{fig:alledt}. A detailed frequency analysis was performed on this
data set, using two different algorithms: the Jurkevich-Stellingwerf {\sc pdm}
method \citep{jurkevich71, stellingwerf78}, and the Lomb-Scargle \citep{lomb76, scargle82}
periodogram. The latter one is essentially a Fourier technique, hence it is more
sensitive for harmonic periodicities. The former technique searches for the
frequency for which the periodogram, devided into a certain number of bins,
shows the smallest spread per bin.

As a first step, a frequency interval has to be defined. A conservative choice
would be the interval between zero and the Nyquist frequency. In our case, when
taking the inverse of the average time gap between neighboring points while
ignoring large gaps, the value of the Nyquist frequency is 44\,c\,d$^{-1}$
(P\,=\,0.55\,h). Such a high frequency will, however, lead to a break-up of
the comet. A rough estimate of the break-up frequency $f_b$ can easily be
calculated by stating that at the equatorial velocity v$_b$ the centrifugal
acceleration equals the gravitational acceleration, and that there is no internal
strength to the body. If one assumes a typical density of 1\,g\,cm$^{-3}$,
then the break-up frequency will be $f_b$\,=\,7.3\,c\,d$^{-1}$
(P$_b$\,=\,3.3\,h). For smaller densities, the break-up frequency will
obviously be smaller; the same is true for aspherical nuclei \citep{jewitt88}.
Here, we adopt a safe upper limit for our frequency interval of
$f$\,=\,20\,c\,d$^{-1}$ (P$_b$\,=\,1.2\,h). The cometary rotation frequencies
that are found up to now \citep{samarasinha04} all fall safely in this range.

\subsection{Non-periodic variability}
As already seen from Fig.~\ref{fig:alledt}, there is a remarkable increase
in brightness in the nights 12-13 and 13-14 January of the observing run. The
brightness increase in those two nights leads to a frequency of
$f$\,=\,0.20\,c\,d$^{-1}$ (P\,=\,5\,d) in the Lomb-Scargle periodograms
(Fig.~\ref{fig:frequencyrslts}), in each filter, and for each of the three
radii (1, 3 and 6 pixels). This frequency is also present in the
Jurkevich-Stellingwerf results, but somewhat less clear.

The increase in brightness is also seen in the ``raw'' magnitudes, i.e. the
magnitudes that are not seeing-corrected. Hence, the brightness increase is
certainly not an artifact of our convolution technique explained in
Sect.~\ref{sect:distancecrrctn}. However, since the convolution technique is
certainly not perfect, it might still be possible that the brightness increase
is related to the seeing values that are somewhat higher than average for these
two nights. We studied this possibilty by making a variability analysis of the
seeing values, and concluded that none of the variability that was found in the
seeing, was still present in the seeing-corrected magnitudes, while the raw
magnitudes showed the same trends as the seeing values, as expected. Hence we
are confident that the seeing effect is adequately accounted for.

Since the total time span of our observational run is only 8 nights, it is not
possible to judge whether this frequency is real (implying that the brightness
increase is a periodic phenomenon), or it is just a unique event. In any case,
since the typical rotational frequencies of cometary nuclei are much higher, it
is unlikely that this frequency is linked to the rotation of the nucleus. The
brightness increase is more likely a non-periodic event, like an outburst.
Possibly it is connected to the sudden change in the solar wind velocity around
that period \citep{degroote08}. It is, however, unlikely that the solar wind
has a direct impact on the comet's nucleus, since it is believed that solar
wind cannot enter the so-called contact surface \citep[$<$10$^3$\,km,][]{ip04}
of the comet.

\subsection{Rotation frequency}\label{subsect:rotfrqnc}
In order to search for an acceptable rotation frequency, the brightness
increase should be removed. This can be done in several ways: e.g. by means of
removing (``prewhitening'') the data with the 5 days periodic component found
earlier. Another possibility is to subtract the mean value of every night. Since
these means are often ill-defined because of the scarcity of the data, we tried
to model the data with a combination of unknown nightly means and a periodic
component. A third possibility is to perform a simultaneous fit of two
frequencies together, one long-term component and one short-term. We will
briefly discuss these three methods and their results.

\subsubsection{Prewithening}\label{subsubsect:prewithnng}
When the frequency $f$\,=\,0.20\,c\,d$^{-1}$ (P\,=\,5\,d) was removed
from the data, we did not find any significant frequencies anymore in the B
filter magnitudes. Two different frequencies pop up in the U and I filter
periodograms. In the U filter, the most significant frequency is present near
$f$\,=\,6.03\,c\,d$^{-1}$ (P\,=\,4.0\,h), but after inspection of the
periodogram, this frequency and some other frequency close to integer values,
seem to be aliases of the frequency 1\,c\,d$^{-1}$. There is little doubt that
this frequency is an artifact, probably introduced by the imperfect extinction
corrections.
Moreover, the probable contribution of OH emission in the U filter (see
Sect.~\ref{sect:observtns}) could explain why this one-day alias is so
prominent for this filter. Indeed, the extinction at OH is very high
($\sim$2\,mag/airmass), so as the comet rises or sets, the signal from OH can
change dramatically. In the I filter, where overall extinction is much less, a
significant frequency is seen at $f$\,=\,2.64\,c\,d$^{-1}$ (P\,=\,9.1\,h) in
both the 1 and 3 pixel size apertures; for the 6 pixel magnitudes, a frequency
of $f$\,=\,3.64\,c\,d$^{-1}$ (P\,=\,6.6\,h), which is a one-day alias of the
former one. The frequency of 2.64\,c\,d$^{-1}$ (P\,=\,9.1\,h) is also found in
this filter and apertures when using the Jurkevich-Stellingwerf method. The
variance reduction by this frequency is another 14\% (from 50\% to 64\%). 

\subsubsection{Free nightly means}
After subtracting the mean values of every night from the data, the results of
the frequency analysis around 2.5\,c\,d$^{-1}$ (P around 9.5\,h) were not
different from the results obtained after prewhitening. However, some spurious,
mostly higher, frequencies are present as well. They appear because the nightly
mean values are not necessarily the true mean values of the light curve, since
the rotation period could be longer than the time span of the comet
observations in one night and since only few observations per night were
available. A solution to this is to perform a harmonic fit through the data,
while taking the mean value for each night as a free parameter. This technique
gave no results for the U filter data, but now a frequency near
$f$\,=\,2.50\,c\,d$^{-1}$ (P\,=\,9.6\,h) appears in the B filter data and
near $f$\,=\,2.59\,c\,d$^{-1}$ (P\,=\,9.3\,h) in the I filter. These results
should be looked at with some caution, since 11 unknowns (three for the
harmonic fit, and one for every night) were to be determined in data sets of
only about 50 observations.

\subsubsection{Two frequencies simultaneously}
Finally we looked also for the best two-frequency fit in the data. We searched
for the best description of the data in the least squares sense with one
long-period component ($f$ between 0.1 and 0.3\,c\,d$^{-1}$, corresponding to
P between 3.3\,d and 10\,d) and a shorter-period component ($f$ between 0.3
and 3.0\,c\,d$^{-1}$, corresponding to P between 8\,h and 3.3\,d). The
$f$\,=\,1.00\,c\,d$^{-1}$ frequency and its aliases are still too dominant in
the U filter data to give reliable results. In the I filter data, a long term
component of 0.18\,c\,d$^{-1}$ (P\,=\,5.6\,d) combined with the frequency
$f$\,=\,2.67\,c\,d$^{-1}$ (P\,=\,9.0\,h) reduces the variance of the data
with 65\%. In the B data the combination of 0.22\,c\,d$^{-1}$ (P\,=\,4.5\,d) and 2.56\,c\,d$^{-1}$ (P\,=\,9.4\,h) describes the observations best
and leads to a  variance reduction of 83\%.

\subsubsection{Final rotation frequency}
\begin{figure}
\resizebox{\hsize}{!}{\includegraphics{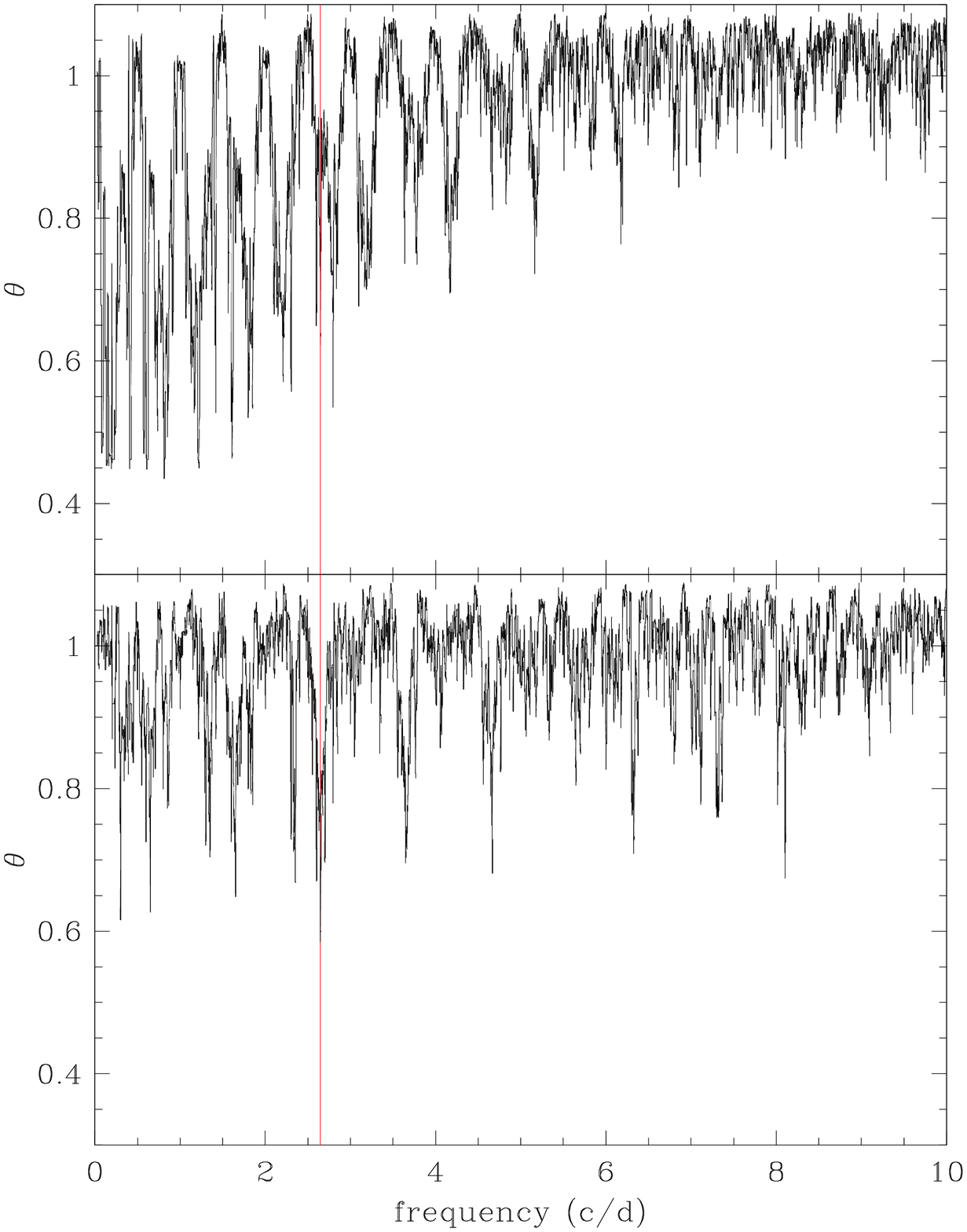}\includegraphics{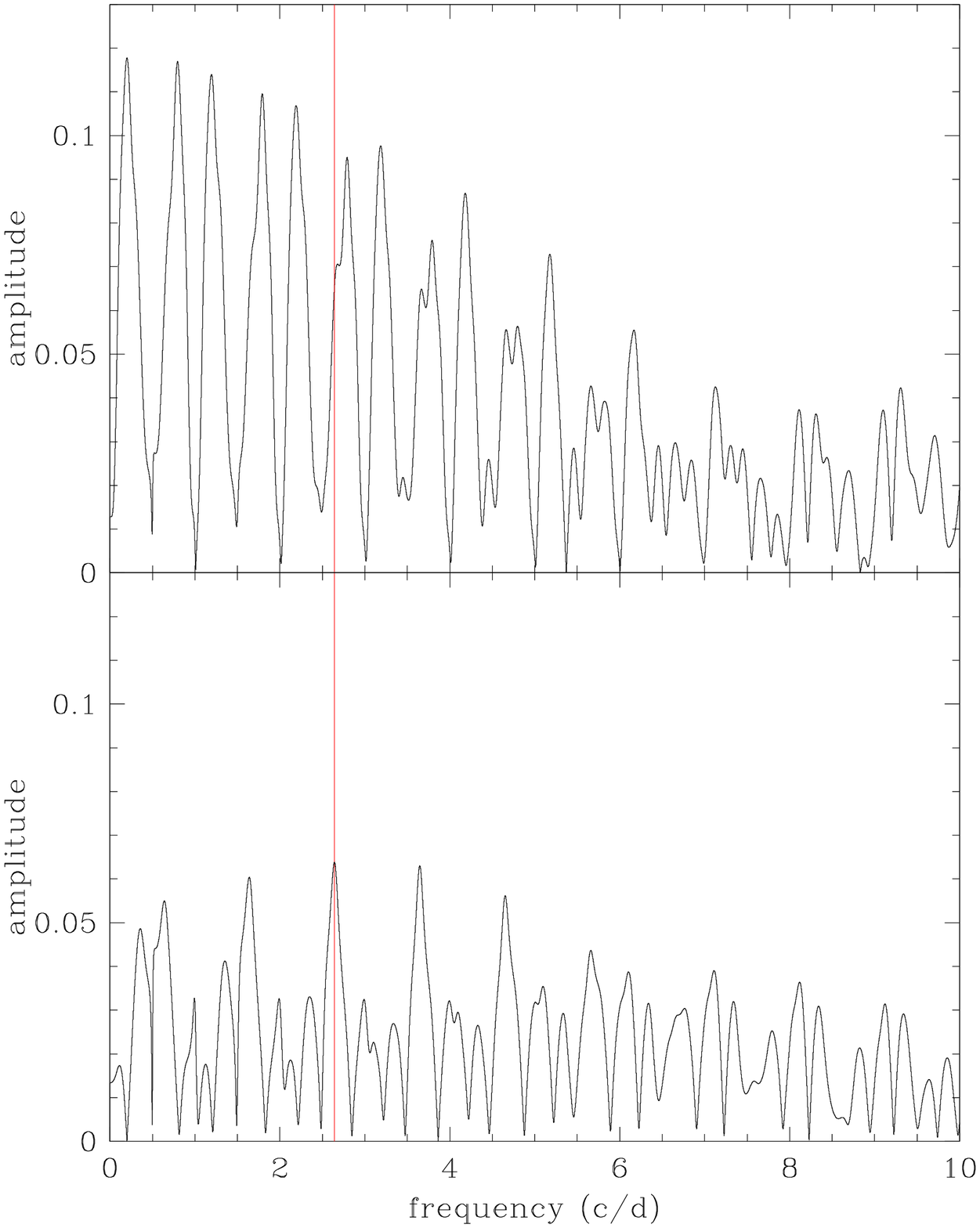}}
\caption{Jurkevich-Stellingwerf {\sc pdm} ({\em left panels})
$\theta$-statistic and Lomb-Scargle ({\em right panels}) periodogram for the I
magnitudes of 1 pixel aperture. In the two {\em upper panels}, the diagrams for
the original data are shown, while in the {\em lower panels}, the diagrams for
the magnitudes prewhitened with $f_1$\,=\,0.20\,c\,d$^{-1}$ (P\,=\,5\,d)
are shown (method {\em \#1} in Table~\ref{tab:scarglerslts}). On each figure,
the frequency $f$\,=\,2.64\,c\,d$^{-1}$ (P\,=\,9.1\,h) is shown with a red line.}\label{fig:frequencyrslts}
\end{figure}

A summary of the frequency analysis is given in Table~\ref{tab:scarglerslts}.
In this table, the results are given in cycles per day (c\,d$^{-1}$), since
this notation allows us to immediately detect one-day aliases, and to compare
directly with the periodograms of Fig.~\ref{fig:frequencyrslts}.
The results of the different methods indicate that a short-term periodicity
is present in the data, with a period between 9.0 and 9.4\,h. Since not all
methods gave consistent results for the B filter data, probably also due to
imperfect extinction correction, we take the value found in the I filter data
(9.1\,h on average) as the best estimate for the rotation period from these
data. The spread of the results given by different techniques, indicates that
the error (1\,$\sigma$) on this value can be as high as 0.25\,h.

\begin{table}
\caption{Summary of the frequency analysis results. Three different methods
were used to get rid of the brightness increase during two nights of our
observational run (12-13 and 13-14 January); see Sect.~\ref{subsect:rotfrqnc}
for more details.}\label{tab:scarglerslts}
\begin{center}
\begin{tabular}{cl}
\hline\hline
\multicolumn{2}{c}{Frequency analysis results}\\
\hline
\multicolumn{2}{l}{\em \#1: after prewithening with $f$\,=\,0.20\,c\,d$^{-1}$}\\
U & strong one day aliases\\
B & no significant $f$\\
I & 2.64\,c\,d$^{-1}$ (1, 3 pix)\\
  & 3.64\,c\,d$^{-1}$ (6 pix)\\
\hline
\multicolumn{2}{l}{\em \#2: after subtracting (free) nightly means}\\
U & no results\\
B & 2.50\,c\,d$^{-1}$ (1, 3 pix)\\
  & 0.50\,c\,d$^{-1}$ (6 pix)\\
I & 2.59\,c\,d$^{-1}$ (all pix)\\
\hline
\multicolumn{2}{l}{\em \#3: two frequency fit (simultaneous)}\\
U & strong one day aliases\\
B & 0.22\,c\,d$^{-1}$ and 2.56\,c\,d$^{-1}$ (all pix)\\
I & 0.18\,c\,d$^{-1}$ and 2.67\,c\,d$^{-1}$ (all pix)\\
\hline
\end{tabular}
\end{center}
\end{table}

\begin{figure}
\resizebox{\hsize}{!}{\includegraphics{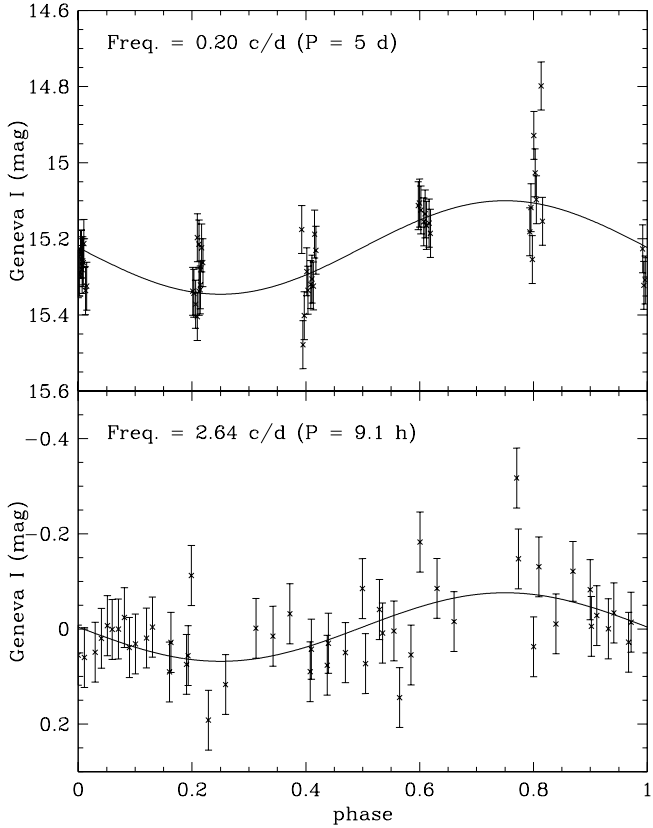}}
\caption{Phase diagrams of the comet's optocentre magnitudes in the I filter
with an aperture radius of 1 pixel. {\em upper panel}: phase plot with
$f_1$\,=\,0.20\,c\,d$^{-1}$ (P\,=\,5\,d); {\em lower panel}: phase plot with
$f_2$=2.64\,c\,d$^{-1}$ (P\,=\,9.1\,h) on the magnitudes prewhitened with
$f_1$ (method {\em \#1} in Table~\ref{tab:scarglerslts}).}
\end{figure}

\subsection{Rotation of Machholz in literature}\label{subsect:rotltrtr}
\citet{sastri05} analysed R band images of C/2004 Q2 (Machholz) taken in the
same period as our observations (January 2005), and report the same period as
ours: P\,=\,9.12\,$\pm$\,1.9\,h ($f$\,=\,2.63\,c\,d$^{-1}$). The period was
derived by studying dust fans visible in these small band images, and adoption
of Fulle's formula \citep{fulle87} between grain velocity and the forces on
each grain. It is a remarkable result that two such different methods lead to
the same value for the rotation period, strengthening of course the confidence
in our results.

\citet{farnham07} observed C/2004 Q2 (Machholz) intensively during several
observing campaigns from January to June 2005, in different filters of the
HB narrowband comet filter system \citep{farnham00}. The paper concentrates
on the CN band images, which reveal two clear jets in roughly opposite
directions relative to the nucleus. The presence of the jets were confirmed by
images taken with the 1-m telescope at the Lulin Observatory, Taiwan, and the
3.6m at ESO, La Silla \citep{lin07}. \citet{farnham07} found that the
morphology of these jets repeated itself on the timescale of hours, and through
an intensive monitoring in April 2005, the authors could phase the jet
morphology on a period of 17.6$\pm$0.05\,h, which is then also identified as
the rotation period of the nucleus. \citet{farnham07} explicitly excludes the
possibility of a period shorter than 17.6\,h, since they did not find similar
morphologies of the CN jets with a time span shorter than this period. Also
their March observations are consistent with this 17.6\,h period. They argue
that it is unlikely that the rotation period of the nucleus should have evolved
from 9.1\,h to 17.6\,h between January and March. Hence, the period of
\citeauthor{sastri05}, as well as the period found by us, is seemingly in
contradiction with the one from \citeauthor{farnham07}. 

There might be, however, a possible explanation that could reconcile the two
results. The method of \citeauthor{farnham07} is based on the visual inspection
of the morphology of the CN jets that emerge from the nucleus. Since it concerns
two opposite jets, there is always one jet dominant in the direction of the
observer. If we assume the jets, or more precise, the regions on the comet's
nucleus where the jets emerge, to be the main contributors in the comet's
variability, then it is obvious that our method catches P/2 and not P. If this
interpretation is correct, the results of \citeauthor{farnham07} are in
agreement with ours and the one of \citeauthor{sastri05} Note that the latter
period was derived by modelling the structure and position of the observed dust
fans, and not by a periodicity search in a series of images, as was done by
\citeauthor{farnham07}.

In the assumption that our method leads to an estimate of P/2 and not P, even
then a small difference persists between our P/2\,=\,9.1\,h and
\citeauthor{farnham07}'s P/2\,=\,8.8\,h. This difference can have different
explanations. Surely, the measurement errors will account for an important part
in this difference, but also another effect can be of importance.
Table~\ref{tab:observationallg} shows a 7\% change in $\Delta$ from the first
to the last night's data.  This produces a commensurate change in the aperture
sizes when measured in km at the comet.  If the variability is caused by
changes in the comet's activity, then the period determination could be
affected by this difference. For example, a burst of activity is seen as a
brightening of the coma. When the leading edge of that burst leaves the
aperture, the light curve has reached its peak and starts fading again. Because
a given aperture is smaller (in km) on earlier nights, the observed peak will
occur earlier than it should relative to the last night.  This produces a
period measurement that is longer than the actual periodicity (defined by the
burst of activity reaching the same distance from the nucleus).  Given the
small pixel dimensions at the comet and the high emission velocities, this
could have a significant effect, and thus may help to explain the difference
between our 9.1\,h and \citeauthor{farnham07}'s 8.8\,h half-periods.

\section{Coma profiles}\label{sect:comaprfls}
The first objective of our observational programme was to study the rotation
of the nucleus through extremely small aperture photometry. The use of the
CCD camera implies, however, that we sample a large part of the inner coma
at each observation. Therefore, we decided to initiate a study of the coma
profile in the three spectral bands at our disposal. In Sect.~\ref{subsect:comaprfls}
we will deduce some physical properties of the coma from these profiles, but
first a small introduction is given of what can be expected theoretically.

\subsection{Theoretical profiles}
Historically, the first quantitative model of a cometary coma was elaborated
by \citet{eddington10}. In \citeauthor{eddington10}'s {\em fountain model} the
comet is assumed to be a uniform and isotropic source of emitting particles
such that their density (e.g., gas or dust) would fall as the inverse square of
the distance from the source, except the emitters are also subject to a uniform
acceleration that pushes on them from a given direction, presumably from
the Sun's direction \citep{combi04}. Continuing \citeauthor{eddington10}'s
work, \citet{wallace58} showed that the column density N in the line of sight,
and hence the brightness $B$ in the case of an optically thin coma, is given by
\begin{displaymath}
N = Q / (4{\rm v} \rho)
\end{displaymath}
where $Q$ is the global particle production rate, v is the initial uniform
outflow speed and $\rho$ is the projected distance on the sky plane of the nucleus.
Generally the logarithmic derivative $m$\,=\,$d\log$B($\rho$)/$d\log$$\rho$ is used
to describe the coma profile, and hence $m$\,=\,$-$1 in \citeauthor{wallace58}'s
equation. Deviations from the $m$\,=\,$-$1 profile are expected at distances
where the solar radiation pressure becomes important. More specifically,
\citet{jewitt87} showed that this distance is of the order of
\begin{equation}
X_R \approx {\rm v}_{\rm gr}^2 r_h^2 / [2 \beta g_{\rm sun}(1)  ] \label{eq:jewittmch}
\end{equation}
where v$_{\rm gr}$ is the terminal ejection velocity of a grain, $g_{\rm sun}$
is the solar gravity at 1\,AU, and $\beta$ is the ratio of the acceleration of
a grain due to radiation pressure to the local solar gravity. Beyond this point
$X_R$, one enters the {\em outer coma} where $|m|$\,$>$\,1. \citet{jewitt87}
derive from both Monte Carlo simulations, as well as from an analytical
derivation, that $m$\,$\simeq$\,$-$1.5 in this region, slightly varying
depending on the phase angle and the radiation pressure.

The theoretical profiles that are discussed here assume that the coma has
constant (though not necessarily isotropic) activity.  In reality,
C/2004 Q2 (Machholz) is known to have significant jet activity
\citep{farnham07}, and jets produce structures that can affect the radial
profiles in the coma.  Jets on rotating nuclei frequently produce curved or
spiral shaped jets. Radial profiles may cross these curved features, causing
the inner or outer region to be brighter than otherwise might be expected. 
This is especially true in the inner coma we are measuring. Since activity
changes are invoked as the cause of the variability that defines the cometary
period (Sect.~\ref{subsect:rotltrtr}), the profiles are expected to be
influenced by this effect.

\subsection{Coma profiles of C/2004 Q2 (Machholz)}\label{subsect:comaprfls}
\begin{figure}
\resizebox{\hsize}{!}{\includegraphics{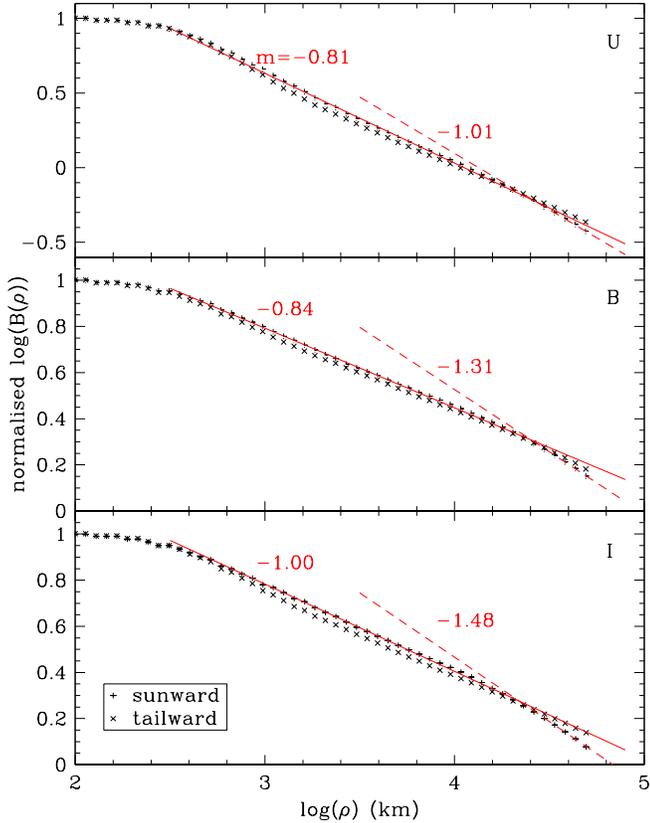}}
\caption{Coma profiles of three representative observations for the three
filters used in our programme. Crosses represent the profile in the sunward
direction, while the $\times$es show the profile in the direction of the ion
tail. The coma in the I filter shows a clean $m$\,=\,$-$1 profile for
$\rho$\,$<$\,10$^{4}$\,km and $m$\,=\,$-$1.5 for $\rho$\,$>$\,10$^{4}$\,km, the
profiles in the U and B filter have smaller values for $|m|$, indicating the
influence of gas emissions in these spectral bands.}\label{fig:resultsprfls}
\end{figure}

To construct representative coma profiles of C/2004 Q2 (Machholz), we took for
each filter one observation taken in the night 08-09 January 2005, which was a
night with optimal conditions and seeing (see Fig.~\ref{fig:robodimmsng}). For
each of these three observations, we constructed two coma profiles, one in the
direction of the sun, and one in the direction of the ion tail. The profiles
were constructed by dividing the coma into concentric annuli, with radii that
are (logarithmically) equidistant. For each of these annuli, two medians
were taken: one of the sunward half of the annulus (sunward profiles in
Fig.~\ref{fig:resultsprfls}) and one of the tailward half of the annulus
(tailward profiles in Fig.~\ref{fig:resultsprfls}). The profiles were
normalised so that the brightness of the optocentre reaches unity, and plotted
on a logarithmic scale. The resulting profiles are shown in
Fig.~\ref{fig:resultsprfls}, and linear fits were made in the two regimes
2.6\,$<$\,$\log \rho$\,$<$\,4 and 4\,$<$\,$\log \rho$ ($\rho$ in km). The values
for $m$ are also given in this figure.

While the profiles in the I are fully compatible with the basic models presented
in the previous section, the profiles in U and B are clearly deviating from
these predictions. From Fig.~\ref{fig:lowrsltn} it is clear that in both the
U and B bands strong emissions are present, while the I filter is relatively
free from strong emission lines. The presence of these emissions breaks
the assumption of conservation of mass of a certain species, since the gas can
dissociate and/or ionise to other species. The development of theoretical
profiles taking these effects into account were initiated by \citet{haser57},
and are specifically valid when using small band filters concentrated on those
emission described in the profile. Since we are using broadband filters in
this study, sampling both continuum and emission features, it is very difficult
to construct a valid profile for the U and B filters. The I filter, however, 
is dominated by the continuum, leading to a clean $m$\,=\,$-$1 profile in the
inner coma.

From the change in $m$ from $-$1 to $-$1.5 at a distance of
$X_R$\,=\,21.6\,$\pm$\,1.3\,$\times$\,10$^4$\,km, we can estimate the terminal
ejection velocity of the grains using formula (\ref{eq:jewittmch}). Typically,
one assumes $\beta$\,$\simeq$\,1, which corresponds at 1\,AU to spherical grains
with a diameter of about 1\,$\mu$m. This way, we can estimate
v$_{\rm gr}$\,=\,1609\,$\pm$\,48\,m\,s$^{-1}$, which is much more than we
expect from the empirical law of Bobrovnikoff-Delsemme \citep{delsemme82}:
v$_{\rm gr}$\,=\,580/$\sqrt{r_h}$\,=\,524\,m\,s$^{-1}$
\citep[e.g.][]{schulz93}. Deviations from this law of this magnitude may be
attributed to stages of high activity. For distances smaller than
$\log \rho$\,$\la$\,2.6, there is also a clear deviation from the $m$\,=\,$-$1
trend. At this scale ($\rho$\,$\la$\,400\,km), other effects come into play. The
acceleration zone close to the nucleus has certainly some influence at this
scale \citep{jewitt87}, but most importantly it is the seeing that wipes
out the $m$\,=\,$-$1 profile close to the optocentre.

\begin{figure}
\resizebox{\hsize}{!}{\includegraphics{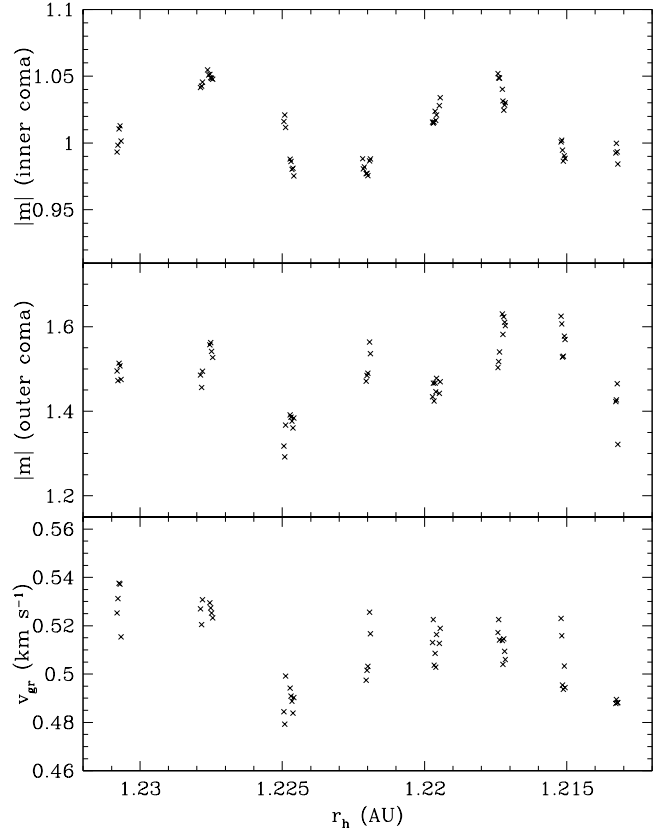}}
\caption{Temporal evolution (in function of the heliocentric distance $r_h$) of
the inner and outer coma profiles represented by their $m$-values, and of the
terminal ejection velocity determined with formula (\ref{eq:jewittmch}). The
terminal ejection velocity is related to the heliocentric distance $r_h$ of the
comet, but the difference of $\Delta r_h$\,$\simeq$\,0.02\,AU between the
beginning and ending of our observational run is too small to establish global
trends.}\label{fig:evolprfls}
\end{figure}

By calculating these $m$-values for all observations, we can search for
temporal variations or global trends in the profiles. We concentrate on the
profiles in the I band because of their agreement with the theoretical
profiles. We did not use observations containing bright background stars for
this purpose, since the profiles of such observations are heavily disturbed
by the star tracks. For the determination of the inner and outer coma profiles,
respectively 56 and 51 observations were used, and the fitted $m$-values are
shown in Fig.~\ref{fig:evolprfls}. From these $m$-values, it is clear that the
profiles show some variation indeed, but overall we can safely state that
the profiles are quite stable throughout our observational run. The spread on
the $m$-value for the inner coma for example, is $m$\,=\,$-$1.01\,$\pm$\,0.03;
for the outer coma this is $m$\,=\,$-$1.49\,$\pm$\,0.08. The seemingly periodic
behaviour of the inner coma profile (Fig.~\ref{fig:evolprfls}, {\em upper
panel}) is very peculiar, as we cannot link it to another phenomenon: not with
the global brightness trend we encountered previously, nor with the deduced
period (P or P/2), nor with any solar wind events. The observed trend is also
noticeable for the inner coma in the U and B band, but not in any of the
$m$-values of the outer coma. The fact that we cannot match the variations in
the slopes to any known periodicity might be explained if the slopes are being
affected by coma structures, as discussed earlier.  If the slopes are affected
by temporary structures, then sparsely sampling the slope at various times may
not reveal variations that change with the rotation period.

On Fig.~\ref{fig:evolprfls} the evolution of the terminal ejection velocity of
the grains v$_{\rm gr}$ is also shown. This quantity is related to the
heliocentric distance $r_h$ of the comet, but the difference for $r_h$ between
the beginning and ending of our observational run is too small to see any
noticeable effect.

\section{Discussion}\label{sect:generaldscssn}
Rotation periods or periodicities are currently known for about 25 comets.
A comprehensive inventory of the currently known or suspected rotation periods
is given in \citet{samarasinha04}, while a more critical list of 15 comets is
given in \citet{jorda00}. The mean of the spin periods given in
\citeauthor{jorda00}'s list is 20\,h, but decreases to 12\,h when constrained
to the short-period comets. Moreover, there seems to be a hint for a relation
between rotation period and the size of the nucleus, with larger nuclei having
longer periods. However, the small number of published periods does certainly
not allow us to draw any final conclusions concerning statistical dependencies.

The difference between the period found by \citet{sastri05} that we confirmed
here, and the period found by \citet{farnham07} is puzzling. \citet{crifo99}
claim that the present circumnuclear models have too uncertain or even
contradictory assumptions, implying that physical properties for the nucleus
derived from observations of jets or fans, can deviate considerably from
reality. However, in the case of C/2004 Q2 (Machholz), the CN jets observed by
\citet{farnham07} are bright, and the rotation is obvious from a simple
inspection of the phased observations
({\tt\small http://www.astro.umd.edu/$\sim$farnham/Machholz/}). The authors
reject a period shorther than 17.6\,h, but, as explained in
Sect.~\ref{subsect:rotltrtr}, there is a possibility that our method picked
up P/2 instead of P. However, a somewhat uncomfortable inconsistency persists:
if the periodicity of 9.1\,h that we found in the light curve, is induced by the
periodicity in the morphology of the CN jets, then it is not clear why the
period is best seen in the I band, and less clear in the B band (see
Table~\ref{tab:scarglerslts}), which contains the CN emission feature
(Fig.~\ref{fig:lowrsltn}). A possible explanation for why the periodic
variations are visible in the I filter, but not in the B filter, which captures
the CN emission, is through the dust production. \citet{farnham07} observed 
full CN spirals in C/2004 Q2 (Machholz), indicating that they remain active
throughout the rotation. Dust production, on the other hand, tends to shut down
when a source passes into darkness. Thus, the brightness in the B filter, which
has a significant signal from CN, may remain fairly constant, while the I
filter, which is dominated by continuum, reflects the dust turning on and off.
In \citet{farnham07} a future paper is announced in which also the observations
of C/2004 Q2 (Machholz) in the other HB narrowband comet filters are to be
discussed. This could hopefully clarify the different results.

It is clear that a light curve analysis of a comet's optocentre is hampered by
several uncertainties. Particularly, the correction for the seeing effect is a
delicate step, and the introduced uncertainties are difficult to quantify.
Our observational setup is, however, unique in the sense that our spatial
resolution of $\sim$48\,km\,pixel$^{-1}$ is much higher than papers using the
same technique of aperture photometry. Therefore, it is not surpirising that
the period is also recovered even when using a relatively large aperture of 6
pixels.

\section{Conclusion}\label{sect:finalcnclsns}
Comet C/2004 Q2 (Machholz) was monitored using the Merope CCD camera on the
Mercator telescope at La Palma, Spain, in January 2005, during its closest
approach to Earth. 170 Images were recorded in three different bands. Our main
goal was to find a rotation period of the nucleus through aperture photometry
using very small apertures. This technique was already succesfully applied to
other bright comets, but it was recently criticised by \citet{licandro00}.
These authors studied the effect of aperture radii smaller than the seeing
disc, and concluded that many periods published in literature that make use of
this technique are contaminated by this {\em seeing effect}. In our analysis of
comet C/2004 Q2 (Machholz), we took this effect into account by degrading our
images to the least favourable seeing. Luckily, since the ambient conditions
during our observational run were very good, only the images with a superb
seeing had to be degraded slightly.

We searched in each filter and three aperture radii (1, 3 and 6 pixels) for
frequencies between 0 and 20\,c\,d$^{-1}$ (i.e. periods down to 1.2\,h), and
this with different algorithms. After taking into account that a long term
trend is in our data
\citep[which is possibly linked to a solar wind event, see][]{degroote08},
a period near 9.1$\pm 0.2$\,h was found. While this rotation period is in
perfect agreement with the one of \citet{sastri05}, it clearly deviates
from the 17.6\,h found by \citet{farnham07}. However, the methodology
of both methods could explain the possibility that our method in fact finds
P/2 and not P. 

Also coma profiles for the inner coma ($\log \rho$\,$<$\,4.6, $\rho$ in km) were
constructed for each photometric band. In the I-band, a clear $m$\,=\,$-$1 for
$\rho$\,$<$\,10$^{4}$\,km and $m$\,=\,$-$1.5 for $\rho$\,$>$\,10$^{4}$\,km was
found, in agreement with a {\em steady state} coma without creation or
destruction of particles. From this regime change, a terminal ejection velocity
of the grains v$_{\rm gr}$\,=\,1609\,m\,s$^{-1}$ was derived. In the other
filters, smaller $m$-values were found, due to the contamination of strong
gas features in those bands, while the I filter is relatively free from those
emissions.

\acknowledgements{The authors are indepted to the referee (T. Farnham) for his
extensive and detailed report that significantly improved the paper. Hans Van
Winckel and the Mercator staff at La Palma are acknowledged for their generous
and continuous support. MR wants to thank Steven Dewitte and Laurent Delobbe of
the Royal Meteorological Institute of Belgium for the opportunity to finish
this paper, and the warm welcome at my new institute. MR acknowledges financial
support from the Fund for Scientific Research - Flanders (Belgium) and from the
Belgian Federal Science Policy Office.}

\bibliographystyle{aa}

\end{document}